\documentclass[11pt, epsfig]{article}
\pdfoutput=1
\usepackage{epsfig, amsmath, amssymb, amsthm, times}
\usepackage{setspace}
\def\theequation{\arabic{section}.\arabic{equation}}
\renewcommand{\theequation}{\thesection.\arabic{equation}}

\newtheorem {thm}{Theorem}[section]
\newtheorem {lem}[thm]{Lemma}

\newtheorem {cor}[thm]{Corollary}

\theoremstyle{defintion}
\newtheorem {df}[thm]{Definition}

\theoremstyle{remark}
\newtheorem{rem}[thm]{Remark}

\theoremstyle{example}
\newtheorem{ex}[thm]{Example}

\theoremstyle{assumption}
\newtheorem{assum}[thm]{Assumption}

\def\pf{{\it Proof.\;}}

\def\Re{{\mathbb R}}

\def\lbl{\label}
\def\be{\begin{equation}}
\def\ee{\end{equation}}
\def\qed{$\square$}
\def\1{{\bf 1}}
\def\one{\mathbf{1_N}}

\def\rank{\mathsf{rank}}

\title{Synchronization of coupled limit cycles}
\author{Georgi S. Medvedev
\thanks{
Department of Mathematics, Drexel University, 3141 Chestnut Street,
Philadelphia, PA 19104, USA {\tt medvedev@drexel.edu} }}
\begin{document}
\maketitle
\begin{abstract}
A unified approach for analyzing synchronization in 
coupled systems of autonomous differential equations
is presented in this work. Through a careful analysis
of the variational equation of the coupled system we establish a
sufficient condition for synchronization in terms of the geometric
properties of the local limit cycles and the coupling operator.
This result applies to a large class of differential equation
models in physics and biology. The stability analysis is complemented
with a discussion of numerical simulations of a compartmental
model of a neuron.
\end{abstract}

\section{Introduction}
\lbl{intro}
\setcounter{equation}{0}
Synchronization of coupled oscillators has been studied extensively
due to its importance in diverse problems in science and engineering
\cite{Blekhman, mmp, pr01,STR03}. Technological applications 
of synchronization include coordination of activity in power, sensor, 
and communication
networks \cite{DB10}; and  control of the groups of mobile
agents \cite{Olfati07,Ren07}. Many experimental systems
exhibit synchrony: electrical circuits \cite{AVR86}, 
coupled lasers \cite{RT94}, and Josephson
junctions \cite{WCS98}, to name a few. In population dynamics, the theory of
synchronization is used to study collective dynamics
\cite{balerini, sumpter,STR03}. Synchronization plays a prominent role
in physiology and in neurophysiology, in particular. It is important
for information processing in the brain \cite{Sin93}, attention,
arousal \cite{UCS99}, and regulation of the sleep-wake cycles
\cite{GR}. In addition, synchronization underlies several 
common neurodegenerative pathologies such as epilepsy \cite{TWB} 
and Parkinson's Disease \cite{LHL}. This list can be continued. 

A large body of mathematical and physical literature is devoted to 
different aspects
 of synchronization in differential equation models. 
For weakly coupled systems, very effective techniques have been developed 
\cite{KU75,KE88,EK91, BMH, CK00, GH07, HI97, LR03, PJ04} 
(see also Chapter 10 in 
\cite{IZH07} for a survey of available methods).  Synchronization in networks
with symmetries has been studied in \cite{SGP03} (see also \cite{GS06}
and references therein). Chaotic synchronization has been a subject of 
intense research \cite{AVR86,ACH97,BR97, BR97a, FY83,Jos00, PC98}.
For strongly coupled systems, a number of studies
explain synchronization in specific physical and 
biological models \cite{AVR86, BH84,Coo09, GH07, MK, steur09, SR86} and for certain
canonical network topologies such as nearest-neighbor coupling on a lattice
\cite{ACH97, hale97}.

Networks arising in applications feature a rich variety of oscillatory
mechanisms and coupling architectures. Therefore, analytical results 
elucidating synchronization in systems under general assumptions on the 
local dynamics and coupling operators are important.
For systems with strong coupling, such results
are rare. The exception is the work by V.~Belykh, I.~Belykh,
and Hasler \cite{BBH04, BBH06}, where sufficient conditions
for synchronization are given in terms of the properties of the 
graph of the network. The pioneering work 
of Afraimovich, Verichev, and Rabinovich \cite{AVR86} already identified
dissipation produced by the coupling operator as a principal
ingredient in a common mechanism of synchronization. Dissipativity 
of the coupling was shown to be responsible for synchronization in many 
systems of coupled differential equations \cite{AVR86, ACH97,hale97,steur09}. It
is also relevant to stability of spatially homogeneous states
in reaction-diffusion systems \cite{hale_infty}. However, the 
dissipation produced by the coupling alone is often not sufficient 
for synchronization. The analysis of forced
Duffing oscillators coupled through position variables in \cite{hale97} 
shows that the intrinsic dissipativity  of the individual (local) 
subsystems is equally important.
In fact, synchronization is achieved by the interplay of the dissipativity
of the coupling and the intrinsic properties of the local systems.
The analysis in \cite{hale97} describes an important mechanism
of synchronization using the
Lyapunov functions that are specific to the Duffing systems coupled 
via a discrete Laplacian. The goal of this 
paper is to study this mechanism under general assumptions on the
local oscillatory dynamics and for a broad class of coupling schemes.
Through a careful analysis
of the variational equation of the coupled system, we derive a
sufficient condition for synchronization in terms of the geometric
properties of the local limit cycles and the coupling operator.
To achieve this, in the vicinity of the periodic solution of the 
coupled system, we construct a moving frame of reference.
After a series of coordinate transformations, we arrive at 
a system of equations that reveals the interplay
of coupling and local dynamics, and shows their
combined  contribution to the
stability of the synchronous solution of the coupled system.

The key step in this analysis is finding a suitable transformation
for the coupling operator in the moving coordinates.
As a by-product, we develop a rigorous reduction
of the coupled system to the system of equations for the phase
variables. In approximate form, this system of phase equations was
obtained in \cite{medvedev10}. 
Similar to Kuramoto's phase reduction for weakly
coupled systems, we expect that these phase equations
will be useful in studies of  physical and biological 
coupled oscillator models in the strong coupling regime.  
For analytical convenience and because the coupled limit cycle
oscillators are common in applications, in this paper we consider
local systems whose dynamics are generated by limit cycles.
Synchronization of chaotic systems as considered in
\cite{AVR86, ACH97,hale97} is admittedly more appealing and
physically less intuitive than synchronization of coupled limit cycles. 
However, from an analytical point of view, the latter problem 
contains many of the ingredients responsible for synchronization
of systems with more complex dynamics. 
For a discussion of how 
the results of the present study
can be extended to chaotic synchronization
and for related extensions for systems with time-dependent
and nonlinear coupling schemes and randomly perturbed local systems,
we refer the interested reader to \cite{medvedev10}.

The paper proceeds as follows. In Section~\ref{section.assumptions}
we list our assumptions and state the main result. Section~\ref{example}
explains the assumptions made in the previous section in the context
of three examples. This is followed by the proof of the main theorem
in Section~\ref{proof}. 

\section{Assumptions and the main result}\lbl{section.assumptions}
\setcounter{equation}{0}
\subsection{The local dynamics}
We start by discussing the assumptions on a local system:
\be\lbl{local}
\mathsf{
\dot x=f(x),\; x:\Re\rightarrow\Re^n
}
\ee
where $\mathsf{f:\Re^n\rightarrow\Re^n}$ is continuous together with partial 
derivatives up to second order.
We assume that $\mathsf{x=u(t)}$ is a periodic solution of (\ref{local}) of
period $1$ with a nonvanishing time derivative $\mathsf{\dot u(t)}\ne 0,\; 
\mathsf{t\in S^1}.$ Denote the corresponding periodic orbit
$\mathsf{\mathcal{O}=\{x=u(t),\; t}\in \mathsf{S^1}:=\Re^1/\mathbb{Z} \}$. 
Near $\mathcal{O}$, one can introduce an orthonormal 
moving coordinate frame (cf. Theorem VI.1.1, \cite{hale_odes}): 
\be\lbl{basis}
\mathsf{ \{ v(\theta), z_1(\theta), z_2(\theta),\dots, z_{n-1}(\theta)\},
\quad v(\theta)={\dot u(\theta)\over \left|\dot u(\theta)\right|},\;
\;\theta\in} \mathsf{S^1}.
\ee
The change of variables 
\be\lbl{localvar}
\mathsf{
x=u(\theta)+Z(\theta)\rho,\; 
Z(\theta)=\mbox{col}(z_1(\theta),\dots, z_{n-1}(\theta))\in\Re^{n\times(n-1)}
}
\ee
in a sufficiently small neighborhood of $\mathcal{O}$, defines 
a smooth transformation 
$\mathsf{x\mapsto (\theta,\rho)\in \mathsf{S^1}\times \Re^{n-1}}$  
\cite{hale_odes}.

\begin{lem}\lbl{lem.polar}
In new coordinates (\ref{localvar}), 
near $\mathcal{O}$ local system (\ref{local})
has the following form
\begin{eqnarray}
\lbl{loctheta}
\mathsf{
\dot\theta} &=& \mathsf{1+a^T(\theta)\rho+O(\left|\rho\right|^2)},\\
\lbl{locrho}
\mathsf{
\dot\rho} &=& \mathsf{A(\theta)\rho+O(\left|\rho\right|^2),
}
\end{eqnarray}
where
\begin{eqnarray}
\lbl{ltheta}
\mathsf{
a^T(\theta)} &=& \mathsf{ 2{v^T(\theta)\over \left|f(u(\theta))\right|}
\left(Df(u(\theta))\right)^sZ(\theta),
}\\
\lbl{Atheta}
\mathsf{
A(\theta)} &=& \mathsf{Z^T(\theta) Df(u(\theta))Z(\theta)-
Z^T(\theta)Z^\prime(\theta),
}
\end{eqnarray}
where $\mathsf{M^s}$ stands for symmetric part of matrix $\mathsf{M}$,
$\mathsf{M^s:=2^{-1}(M+M^T)}$.
\end{lem}

\noindent
{\bf Notational convention.}
{\it 
 To simplify notation, in the calculations through out this paper, we will
often suppress $\theta$ and $\mathsf{u(\theta)}$ as arguments of $\mathsf{f, Z, v,}$
etc, when the expression of the argument is clear from the context. We continue
to denote the differentiation with respect to $t$ and $\theta$ by dot and prime respectively.
}

\noindent
\pf The proof follows the lines of the proof of Theorem VI.1.1 \cite{hale_odes}.
By plugging (\ref{localvar}) into (\ref{local}), we have
\be\lbl{localvarinloc}
\mathsf{
\left[u^\prime(\theta) +Z^\prime(\theta)\rho\right]\dot\theta +Z(\theta)\dot\rho=
f(u(\theta))+Df(u(\theta))Z(\theta)\rho+Q_1(\theta,\rho),
}
\ee
where
$$
\mathsf{
Q_1(\theta,\rho)=f(u(\theta)+Z(\theta)\rho)-f(u(\theta))-Df(u(\theta))Z(\theta)\rho=
O(\left|\rho\right|^2).
}
$$
By multiplying both sides of (\ref{localvarinloc}) 
by $\mathsf{v^T|f|^{-1}}$, we obtain
\be\lbl{thetadot}
\mathsf{
\dot \theta(1+|f|^{-1}v^T Z^\prime\rho)=1+
|f|^{-1}v^T (Df)Z\rho +O(|\rho|^2).
}
\ee
For small $\mathsf{|\rho|}$, (\ref{thetadot}) can be rewritten
as
\be\lbl{thetainterm}
\mathsf{
\dot \theta=1+ |f|^{-1}v^T \left((Df)Z -Z^\prime\right)\rho 
+O(|\rho|^2).
}
\ee
By differentiating both sides of $\mathsf{v^T(\theta)Z(\theta)=0}$,
we have
\be\lbl{byparts}
\mathsf{
v^TZ^\prime=-(v^\prime)^TZ= -v^T(Df)^TZ.
}
\ee
By plugging in (\ref{byparts}) into (\ref{thetainterm}), we 
arrive at (\ref{ltheta}). Next, we multiply (\ref{localvarinloc}) 
by $\mathsf{Z^T}$, and  using $\mathsf{\dot\theta=1+O(|\rho|)},$ 
we derive (\ref{locrho}).\\
\qed

Next we formulate our assumption on the stability of the local limit cycle.
\begin{assum}\lbl{localcycle}
Let $\mu_1(\theta)$ denote the largest eigenvalue of $\mathsf{A^s(\theta)}$
(cf. (\ref{Atheta})).
We assume 
\be\lbl{unistable}
\int_0^1 \mu_1(u)du=:-\mu<0.
\ee
\end{assum}
Assumption~\ref{localcycle} implies exponential stability
of the trivial solution of the linear periodic system
\be\lbl{persyst}
\dot\rho =\mathsf{A}(t)\rho.
\ee
For convenience of the future reference, we formulate this 
statement as a lemma.

\begin{lem}\lbl{expstab} 
Suppose $\mathsf{A}(t)$ is a continuous 
periodic matrix of period $1$.
Then (\ref{unistable}) implies that for any $0<\varepsilon<\mu$,
\be\lbl{estfund}
|R(t)R^{-1}(s)|\le C_1\exp\{-(\mu-\varepsilon)(t-s)\}, \; t\ge s,
\ee
where $C_1>0$ and $R(t)$ stands for the principal matrix solution 
of (\ref{persyst}) \cite{hale_odes}.
\end{lem}

\noindent \pf Let
\be\lbl{Lyapexp}
\nu(t):={1\over t} \int_0^t \mu_1(s)ds.
\ee
By periodicity of $\mu_1(\theta)$ and (\ref{unistable}),
\be\lbl{limnu}
\lim_{t\to\infty} \nu(t)=-\mu<0.
\ee
Further,
$$
{d\over dt} |\rho(t)|^2= 
2\rho^T\mathsf{A^s}(t)\rho\le 2\mu_1(t)|\rho(t)|^2
$$
and, by Gronwall's inequality,
\be\lbl{Vazh}
|\rho(t)|\le |\rho(0)| \exp\{\nu(t)t\}.
\ee
This in turn implies that for large $t\gg 1$ and 
arbitrary $0<\varepsilon<-\mu$
\be\lbl{asympt}
{ |R(t)\rho(0)|\over |\rho(0)|} \le \exp\{-(\mu-\varepsilon)t\}.
\ee 
On the other hand, by the Floquet theorem,
\be\lbl{Floc}
R(t)=P(t)\exp\{t\mathsf{\tilde A}\},
\ee
where $P(t)$ and $\mathsf{\tilde A}$ are periodic and constant
matrices respectively. By comparing (\ref{asympt}) and (\ref{Floc}),
we conclude that all eigenvalues of $\mathsf{\tilde A}$ must have
negative real parts. This yields (\ref{estfund}).\\
\qed

\subsection{The coupling operator}
 By the coupled system, we call a collection of
$N$ local dynamical systems (\ref{local}) interacting via a 
linear coupling operator $D:\Re^{Nn}\to\Re^{Nn}$
\be\lbl{coupled}
\dot x=f(x)+g Dx,\; 
x=\left(\mathsf{x^{(1)},x^{(2)},\dots,x^{(N)}}\right)\in\Re^{Nn},
\ee
where $f(x)=\left(\mathsf{f(x^{(1)}),f(x^{(2)}),\dots, f(x^{(N)})} \right)\in\Re^{Nn}$
and  $g\ge 0$ is a parameter controlling the strength
of interactions. 
In this paper, we consider separable schemes \cite{medvedev10}:
\be\lbl{separable}
D=\mathbf{D}\otimes\mathsf{L},
\ee
where $\mathsf{L}\in\Re^{n\times n}$, $\mathbf{D}\in\Re^{N\times N}$,
and $\otimes$ stands for the Kronecker product \cite{harville}.
From the modeling point of view, separable coupling is natural as 
it reflects two levels of the network organization. The global
network architecture of interconnections between local systems
is captured by $\mathbf{D}$. Matrix $\mathsf{L}$ reflects the
organization of the coupling at the level of a local system:
it shows what combination of local variables participates
in the coupling. The separable structure of the coupling 
translates naturally into the stability analysis of
the synchronous solution, revealing what features 
of the global and local organization of the coupling are important
for stability.
\begin{df}
By synchronous periodic solution of (\ref{coupled})
(when it exists) we call 
\be\lbl{sync_solution}
x=u(t):=\one\otimes\mathsf{u(t)},\;\;\one:=(1,1,\dots,1)^T\in \Re^N,
\ee 
where $\mathsf{u(t)}$ is a periodic solution of the local
systems (\ref{local}).
\end{df}
Next we specify the structure of the separable coupling operator
$D$ (cf. (\ref{separable})). There are two (sets of) assumptions.
The first simpler assumption ensures that the coupled system admits
a synchronous solution. The second set of assumptions guarantees 
the stability. As far as existence is concerned, we need to postulate
that $\one\in\ker(\mathbf{D})$. We further assume that 
$\ker(\mathbf{D})$ is one-dimensional to limit our study to connected
networks. Thus, we assume 
\be\lbl{kerD}
\mathbf{D}\in\mathcal{K}=\left\{\mathbf{M}\in\Re^{N\times N}:~
\ker~\mathbf{M}=\mbox{Span}~\{\mathbf{1_N} \}\right\}.
\ee 

Next, we turn to assumptions that ensure stability of the 
synchronous solution. To this end, we define an $(N-1)\times N$ matrix
\be\lbl{defineS}
\mathbf{S}=
\left(\begin{array}{cccccc}
-1 & 1 & 0& \dots &0& 0 \\
0 & -1& 1 & \dots& 0& 0\\
\dots&\dots&\dots&\dots&\dots&\dots\\
0 &0 & 0& \dots &-1 & 1
\end{array}
\right).
\ee
In the stability analysis of the synchronous solution, we will use
matrix $\mathbf{\hat D}\in \Re^{(N-1)\times (N-1)}$ defined by the following
relation
\be\lbl{hatD}
\mathbf{SD}=\mathbf{\hat DS}
\ee
For any $\mathbf{D}\in\mathcal{K}$, $\mathbf{\hat D}$ is well-defined 
(cf. \cite{medvedev10b}, see also Appendix in \cite{medvedev09}). 
The spectrum of 
$\mathbf{\hat D}$ is important for synchronization. This motivates
the following definition.
\begin{df}\lbl{dissipative}
Matrices from
\be\lbl{diss}
\mathcal{D}=\left\{\mathbf{M}\in\mathcal{K}:\; \mathbf{x^T \hat Mx} <0\; 
\forall \mathbf{x}\in\Re^{N-1}/\{0\}
\right\}
\ee
are called dissipative.
\end{df} 

Finally, we state our assumptions on matrix $\mathsf{L}$
describing the local organization of the coupling .
\begin{assum} $\mathsf{L}$ is (symmetric)  positive
semidefinite, i.e., $\mathsf{L^T=L}$ and $\mathsf{x^TLx\ge 0}$
$\mathsf{\forall x\in\Re^n}$. 
\end{assum}
The following definition distinguishes two important cases that 
come up in the stability analysis of the synchronous solutions
of (\ref{coupled}) and (\ref{separable}).
\begin{df}\lbl{rank}
If $\mathsf{L}$ is positive definite,
we say that the coupling is full (rank), otherwise we call it
partial (rank).
\end{df}
\begin{rem} In a slightly different form, the full and the partial 
coupling schemes were introduced in \cite{hale97}.
\end{rem}

Dissipative matrices yield synchronization when the  
coupling is full and sufficiently strong. Synchronization in the partially 
coupled systems requires an additional hypothesis.

\begin{assum}\lbl{assumptions}
If $\dim\ker~\mathsf{L}=l>0$, let $\{\mathsf{p_1,p_2,\dots, p_l}\}$ be
an orthonormal basis of $\ker~\mathsf{L}$. Denote orthogonal matrix 
$\mathsf{O}=\mathsf{col}(\mathsf{v,z_1, z_2,\dots, z_{n-1}})$ and 
matrix
\be\lbl{B1s}
\mathsf{L_1^s(t)}=
\left(\begin{array}{cc} {1\over 2}c(t)  & \mathsf{v^T(t)(Df(u(t)))^sZ(t)} \\ 
                         \mathsf{Z^T(t)(Df(u(t)))^sv(t)} & 
\mathsf{A^s(t)}-{1\over 2}c(t)\mathsf{I_{n-1}} 
\end{array}\right),
\ee
where $c=\mathsf{v^T(Df)v}$.
Define matrix 
\be\lbl{matrixG}
\mathsf{G}(t)=(\mathsf{g_{ij}}(t))\in\Re^{l\times l},\; 
\mathsf{g_{ij}}(t)=(\mathsf{L_1^s}(t)\mathsf{O^T}(t)\mathsf{p_i, O^T}(t)\mathsf{p_j}).
\ee
Let $\mathsf{\lambda_1(t)}$ denote the largest eigenvalue of 
symmetric matrix $\mathsf{G(t)}$. We assume
that
\be\lbl{intlambda}
\int_0^1\mathsf{\lambda_1(t)dt} <0.
\ee
\end{assum}

Assumption~\ref{assumptions} captures the geometric properties of the local limit 
cycle through matrix $\mathsf{L_1^s}$ (cf. (\ref{B1s})), which depends on the 
basis functions $\{\mathsf{v,z_1, z_2,\dots, z_{n-1}}\}$ defined in the vicinity
of the limit cycle and the Jacobian $\mathsf{Df(u(t))}$ evaluated
along the periodic orbit and the properties of the coupling operator through
the basis for the kernel of $\mathsf{L}$, $\{\mathsf{p_1,p_2,\dots, p_l}\}$,
entering the definition of $\mathsf{G}$ (cf. (\ref{matrixG})).
Condition (\ref{intlambda}), therefore, reflects the interplay of the 
intrinsic properties of  the limit cycle and the coupling operator
and their contribution to the stability of the periodic solution
of the coupled system.
Having reviewed the assumptions, we state the main result of this
paper.

\begin{thm}\lbl{thm.full}
Suppose a periodic solution of local system (\ref{local}) is stable 
in the sense of Assumption~\ref{localcycle} (cf. (\ref{unistable})) 
and the coupling operator in (\ref{coupled}) is
separable (cf. (\ref{separable})) and such that $\mathbf{D}\in\mathcal{D}$
and $\mathsf{L}$ is positive semidefinite.
If the coupling is partial, in addition, we assume that (\ref{intlambda})
holds. Then for sufficiently large $g>0$, the coupled system (\ref{coupled})
has an exponentially stable synchronous limit cycle.
\end{thm}

\section{Discussion and examples}\lbl{example}
\setcounter{equation}{0}

We precede the proof of Theorem~\ref{thm.full} with a discussion of the 
assumptions and the implications of the theorem. We illustrate our techniques 
with three examples. In the first example, we use a planar vector
for which Assumption~\ref{assumptions} can be checked by a simple 
explicit calculation. 
The second example is used to explain the assumption 
that the coupling operator $\mathbf{D}$ is dissipative and to elucidate 
the range of networks covered by this assumption. The third example is 
meant to illuminate the distinction between the full and partial coupling 
schemes in the context of a biophysical model of a neuron, and to show 
how one verifies the assumptions of the Theorem~\ref{thm.full} 
in practice. 

\subsection{Coupled radially symmetric oscillators}
Consider radially symmetric local vector 
field:
\be\lbl{rad}
\mathsf{ \dot x=f(x),\;\;
f(x)=
}
\begin{pmatrix} 
\mathsf{
x_1-x_2-x_1(x_1^2+x_2^2)}\\
\mathsf{
x_1+x_2-x_1(x_1^2+x_2^2),
}
\end{pmatrix}\;\;
\mathsf{x=(x_1,x_2)^T\in\Re^2}.
\ee
For coupled local systems (\ref{rad}) Theorem~\ref{thm.full} yields 
the following sufficient condition for 
synchronization.
\begin{cor}\lbl{uni}
Coupled system (\ref{coupled}) with separable coupling (\ref{separable}) 
and  local vector field (\ref{rad}) has an exponentially stable synchronous
limit cycle if~ $\mathbf{D}\in\mathcal{D}$ and
nonzero matrix $\mathsf{L}$ is positive semidefinite.
\end{cor}
\pf
If $\mathsf{L}$ is positive definite then Corrolary~\ref{uni}
follows from Theorem~\ref{thm.full}. Suppose
that $\mathsf{L}$ has a $1D$ kernel spanned by
$$
\mathsf{p=(p_1,p_2)^T}.
$$
Let
$$
\mathsf{u(t)}=(\cos t, \sin t)^T,\; \mathsf{v(t)}=(-\sin t,\cos t)^T,\;
\mathsf{z(t)}=(-\cos t, -\sin t)^T.
$$
Then
$$
\mathsf{
O=}
\begin{pmatrix} \cos t & -\sin t \\ 
\sin t &  \cos t
\end{pmatrix} \quad \mbox{and}\quad
\mathsf{L_1^s=}
\begin{pmatrix} 0 & 0 \\ 
\mathsf 0 & -2 
\end{pmatrix}.
$$
Further, $\mathsf{ q=O^Tp}$ and
$$
\mathsf{
\lambda_1(t)=q(t)^T L_1^sq(t)}=-2(\mathsf{p_1}\cos t+\mathsf{p_2}\sin t)^2
$$
and
$$
\mathsf{
 \int_0^{2\pi}\lambda_1(t)dt= -(p_1^2+p_2^2)<0.
}
$$
\qed
\begin{ex} Let $\mathsf{f=(f_1,f_2)^T}$ be as in (\ref{rad}) and
consider
\begin{eqnarray*}
\dot x_1^{(i)} &= & \mathsf{f_1}(x_1^{(i)},x_2^{(i)})
+\sum_{j=1}^N h_{ij} (x_1^{(j)}-x_1^{(i)}),\\
\dot x_2^{(i)} &= & \mathsf{f_2}(x_1^{(i)},x_2^{(i)}),
\quad i=1,2,\dots, N.
\end{eqnarray*}
In this example,
$$
\mathsf{L}=\begin{pmatrix} 1 & 0\\ 0 & 0\end{pmatrix}
$$
is positive semidefinite. Suppose 
$h_{ij}=h_{ji}\ge 0$ and denote
$$
\mathbf{D}=(\mathbf{d_{ij}}), \;\;
\mathbf{d_{ij}}=\left\{ \begin{array}{cc}
h_{ij}, & i\neq j,\\
-\sum_{k\neq i} h_{ik}, & i=j.
\end{array}
\right.
$$
If $\dim\ker \mathbf{D}=1$ then $\mathbf{D}\in\mathcal{D}$,
by Gershgorin's Theorem. In general, $\mathbf{D}\in\mathcal{D}$
does not have to be symmetric.
For a more complete description of dissipative matrix we refer
the reader to \cite{medvedev10b} (see also Theorem~\ref{QLthm}
below).
\end{ex}

\subsection{The consensus protocol}\lbl{consensus}
To elucidate what features of the coupling are important for synchronization
we choose consensus protocols, 
a framework used for modeling coordination in the groups of dynamic agents
\cite{Ren07}. 
The continuous time variant for this problem deals with
$N$ agents whose states are given by $x^{(i)}(t)$, $i=1,2,\dots,N$ and are 
governed by the following system of differential equations:
\be\lbl{ConsEqn}
\dot x^{(i)}=\sum_{j=1}^N d_{ij} (x^{(j)}-x^{(i)}),\; i=1,2,\dots,N.
\ee
Here, weights $d_{ij}, i\ne j$ (which for simplicity we take constant) 
describe the interactions between two distinct 
agents $x^{(i)}$ and $x^{(j)}$. After setting $d_{ii}=-\sum_{j\ne i} d_{ij}$, we 
rewrite (\ref{ConsEqn}) in a vector form
\be\lbl{ConsSys}
\dot x = D x, \;D=(d_{ij})\in\mathcal{K},\; x=(x^{(1)},x^{(2)},\dots, x^{(N)}). 
\ee
The problem data can be conveniently represented
by a weighted graph $\mathcal{G}=(\mathcal{V,E,} \{d_{ij}\})$, 
where vertex set $\mathcal{V}=\{1,2,\dots,N\}$
lists all agents, the pairs of interacting agents are recorded in 
the edge set $\mathcal{E}$, and the weights $\{ d_{ij}\}$ quantify 
the intensity of interactions.
If $\mathcal{G}$ is connected $\ker~D=\mbox{Span}\{\one\}$.
Both positive and negative weights $\{ d_{ij}\}$ (corresponding to 
synergistic and antagonistic interactions respectively) are admissible.
Likewise,
the interactions do not have to be symmetric, i.e., 
$d_{ij}$ may differ from $d_{ji}$.
In designing consensus protocols, one is interested in weighted graphs  
yielding convergence to spatially homogeneous state
\be\lbl{ConsAsympt}
|x^{(i)}(t) - x^{(j)}(t)|\to 0 \;\mbox{as}\; t\to\infty.
\ee
The following questions related to (\ref{ConsSys}) are important
in applications: determining 
the rate of convergence in (\ref{ConsAsympt}) and relating 
it to the network topology and weight distribution \cite{Boyd03, Leonard10},
finding an optimal weight distribution yielding 
(\ref{ConsSys}) a desired property such as the fastest convergence 
(under certain constraints) 
\cite{Boyd03,Boyd06} or minimal effective resistance \cite{Boyd08},
or robustness to noise \cite{Leonard10},
to name a few. For studying these questions it is important to describe all
weighted graphs that endow (\ref{ConsAsympt}) with synchronous dynamics.
For the discrete time counterpart of (\ref{ConsEqn}), this question was
answered in \cite{Boyd03}.  The following theorem describes all such graphs
for the continuous time problem (\ref{ConsEqn}).
\begin{thm}\lbl{consensus}
Let $x(t)$ be a solution of the initial value problem (\ref{ConsSys}) with 
initial condition $x_0\in\Re^N$. Then (\ref{ConsAsympt}) holds for any 
$x_0\in\Re^N$ iff $D\in\mathcal{D}$.
\end{thm}

\noindent \pf By multiplying both sides of (\ref{ConsSys}) by $\mathbf{S}$
(cf. (\ref{defineS})), we have
\be\lbl{Eqnfory}
\dot y =\hat D y, \quad y:=\mathbf{S}x.
\ee
The equilibrium of (\ref{Eqnfory}) is asymptotically stable iff the symmetric
part of $\hat D$ is negative definite, i.e., when $D\in\mathcal{D}$.\\
\qed

\begin{figure}
\begin{center}
{\bf a}\epsfig{figure=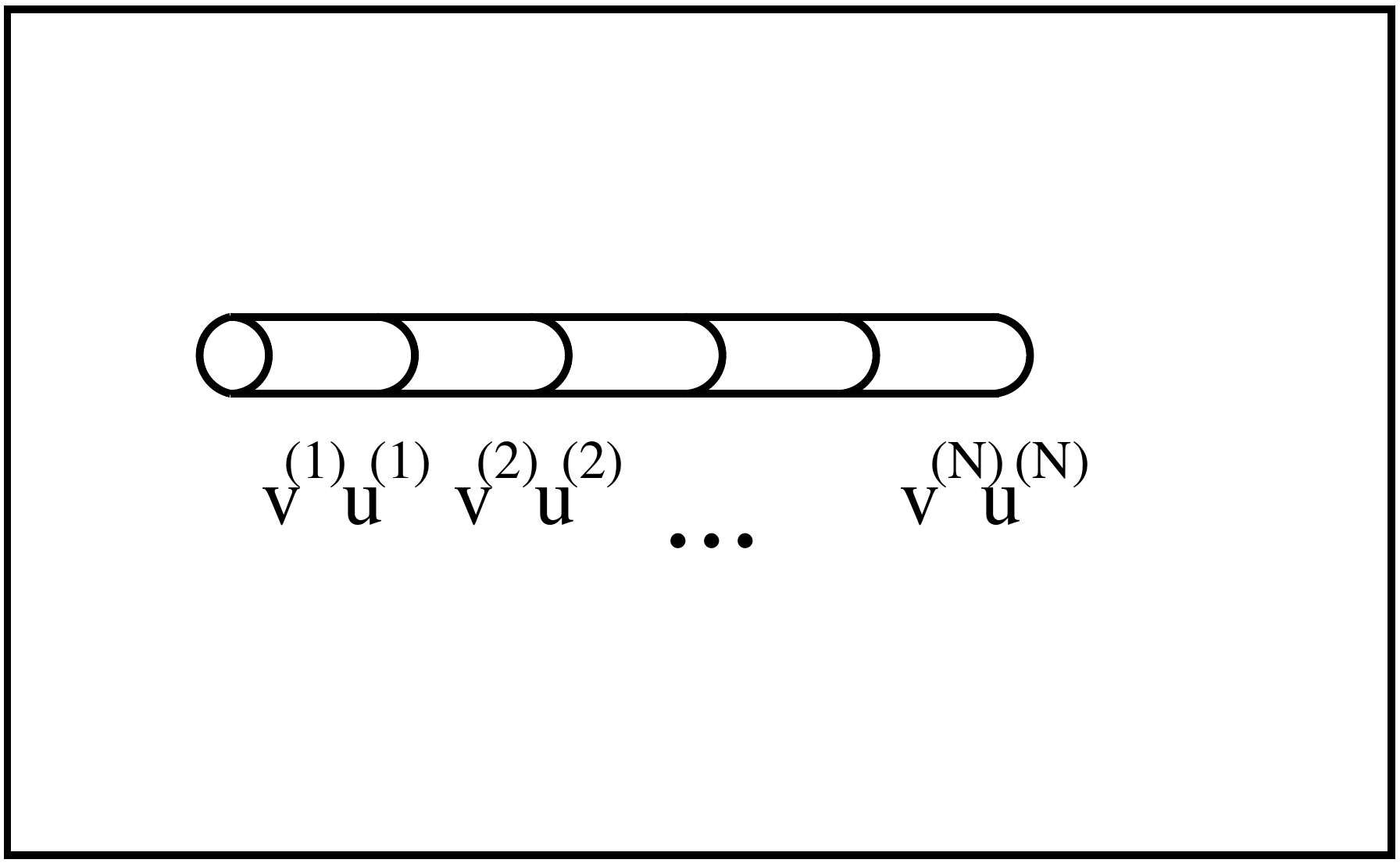, height=2.0in, width=2.5in, angle=0}
\qquad
{\bf b}\epsfig{figure=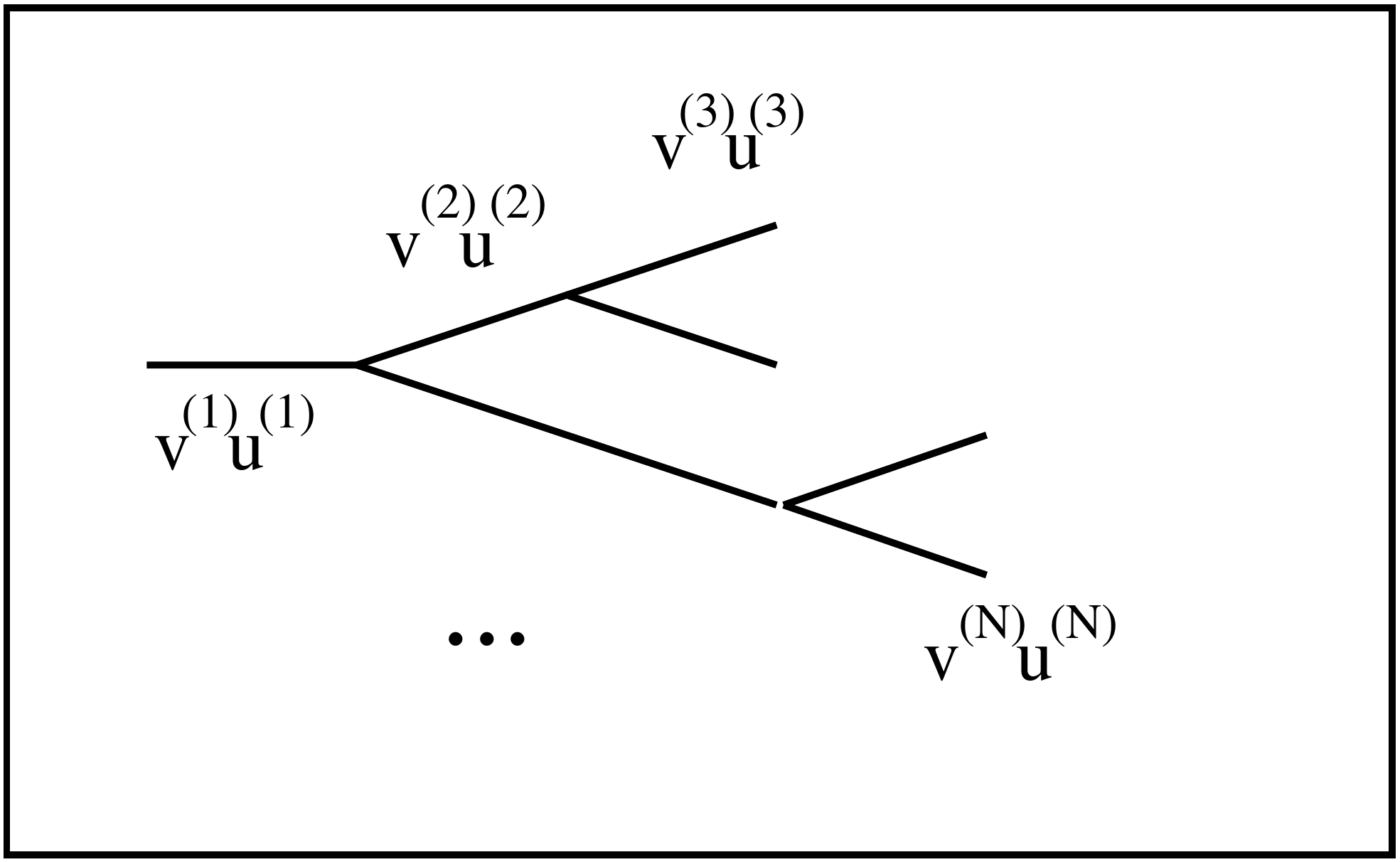, height=2.0in, width=2.5in, angle=0}
\end{center}
\caption{Schematic representation of spatial structure of a 
compartmental model: (a) linear cable, (b) branched cable. 
Dynamic variables $v^{(i)}$ and $u^{(i)}$, $i=1,2,\dots,N$, 
approximate voltage and calcium concentration in each compartment.
}\label{f.1}
\end{figure}

Therefore, dissipative coupling matrices are precisely those that 
enforce synchrony in (\ref{ConsSys}). 
Remarkably, dissipative matrices admit an explicit characterization.
\begin{thm}\lbl{QLthm}\cite{medvedev10}
$\mathbf{D}\in\mathcal{D}$ iff 
\be\lbl{QLambda}
\mathbf{D=Q\Lambda_0}
\ee
for some $\mathbf{Q}\in\Re^{N\times N}$ with negative definite symmetric part
and 
\be\lbl{Lambda0}
\mathbf{\Lambda_0}:=\mathbf{S^TS}=
\left(\begin{array}{cccccc}
1 & -1 & 0 & \dots & 0 & 0\\
-1& 2& -1& \dots& 0& 0\\
\dots& \dots& \dots& \dots &\dots& \dots \\
0& 0& 0& \dots& -1& 1
\end{array}
\right).
\ee
\end{thm}

Theorem~\ref{QLthm} gives a convenient computational formula for 
$\mathbf{\hat D}$:
\be\lbl{compute}
\mathbf{\hat D}=\mathbf{SQS^T}.
\ee
This formula can be used for studying the rate of convergence
of solutions of (\ref{ConsEqn}) to the homogeneous state 
for different network topologies. For some canonical network 
architectures, including nearest neighbor and all-to-all
coupling schemes, the rate of convergence (i.e., the largest
eigenvalue of $\mathbf{\hat D}$) can be found analytically
for other, such as networks with random connection weights, 
the rates can be computed numerically using (\ref{compute}).
We refer an interested reader to \cite{medvedev10, medvedev10b}, 
where these examples are discussed in detail. 

Theorem~\ref{thm.full} extends the argument used in Theorem~\ref{consensus}
to networks whose local systems have multidimensional phase space
and nontrivial dynamics. In the analysis of the general case, in addition
to the global network architecture (i.e. $D\in\mathcal{D}$), the coupling 
organization on the level of the local systems becomes important.
For instance, it matters what local variables and in what manner are
engaged in the coupling. This information is contained in matrix
$\mathsf{L}$ (cf. (\ref{separable})). The analysis of the separable
coupling schemes highlights the importance of the full versus partial
coupling distinction. In the latter case, synchronization depends on the
combination of the properties of the local dynamics and the coupling
operator (cf. Assumption~\ref{assumptions}). 
The example in the following subsection is chosen to 
provide a reader with an intuition for the mechanisms of synchronization
in fully and partially coupled systems.

\subsection{The compartmental model}\lbl{compartment}
The following systems of differential equations is a nondimensional model 
of the dopamine neuron (cf. \cite{MC04}):
\begin{eqnarray}\lbl{da1}
\epsilon \dot v^{(i)}  &=&  g_1(v^{(i)})\left(E_1-v^{(i)}\right)+
g_2(u^{(i)})\left(E_2-v^{(i)}\right)
+\bar g_3\left(E_3-v^{(i)}\right) 
+I_v^{(i)},\\ 
\dot u^{(i)}&=&\omega\left( g_1(v^{(i)})
\left(E_1-v^{(i)}\right)-{u^{(i)}\over \tau}\right)
+I_u^{(i)}, \quad i=1,2,\dots,N.
\lbl{da2}
\end{eqnarray}
Here, $v^{(i)}$ and $u^{(i)}$ approximate
membrane potential and calcium concentration in Compartment $i$ of an axon or a
dendrite of a neuron. Equations (\ref{da1}) and (\ref{da2}) describe the dynamics
in each compartment using Hodgkin-Huxley formalism 
(see \cite{DA} for more background on compartmental models).
The nonlinear functions $g_1(v)$ and $g_2(u)$ and the values of the parameters appearing on the
right hand sides of (\ref{da1}) and (\ref{da2}) are given in the appendix to this paper. Equations (\ref{da1}) and (\ref{da2}) reflect the contribution of the 
principal ionic currents (calcium and calcium dependent potassium currents) to the
dynamics of $v^{(i)}$ and $u^{(i)}$. In addition, the coupling terms
\begin{eqnarray}\lbl{da3}
I_v^{(i)} &=&
g\sum_{j=0}^N {\mathbf d_{ij}} (v^{(j)}-v^{(i)}),\\
\lbl{da4}
I_u^{(i)} &=&
\delta\sum_{j=0}^N {\mathbf d_{ij}} (u^{(j)}-v^{(i)})
\end{eqnarray}
model the electrical current and calcium diffusion between the adjacent 
compartments. The off-diagonal entry $g\mathbf{d_{ij}}$ of matrix 
$g\mathbf{D}$ corresponds to the conductance between Compartments $i$ and $j$.
The structure of the coupling matrix $\mathbf{D}$, i.e., the pattern in 
which nonzero entries
appear in $\mathbf{D}$, reflects the geometry of the neuron. In the simplest case
of a uniform linear cable with no-flux boundary conditions (see Fig.~\ref{f.1}a),
the coupling matrix $\mathbf{D}=-\mathbf{\Lambda_0}$ (cf. (\ref{Lambda0})).
$\mathbf{D}$ may have a more interesting structure, e.g., in the models of dendrites
with more complex spatial geometry (see Fig.~\ref{f.1}b).
As follows from (\ref{da1}) and (\ref{da2}) the coupling is separable with 
\be\lbl{da5}
\mathsf{L}=
\left(
\begin{array}{cc} 1 & 0\\
                  0 & \delta_1 \end{array}
\right). 
\ee
where $\delta_1=g^{-1}\delta$. If $\delta>0$  the coupling is full. If calcium
diffusion is ignored ($\delta=0$) the coupling becomes partial. Below, we discuss
the assumptions of Theorem~\ref{thm.full} in relation to the model at hand.
\begin{figure}
\begin{center}
{\bf a}\epsfig{figure=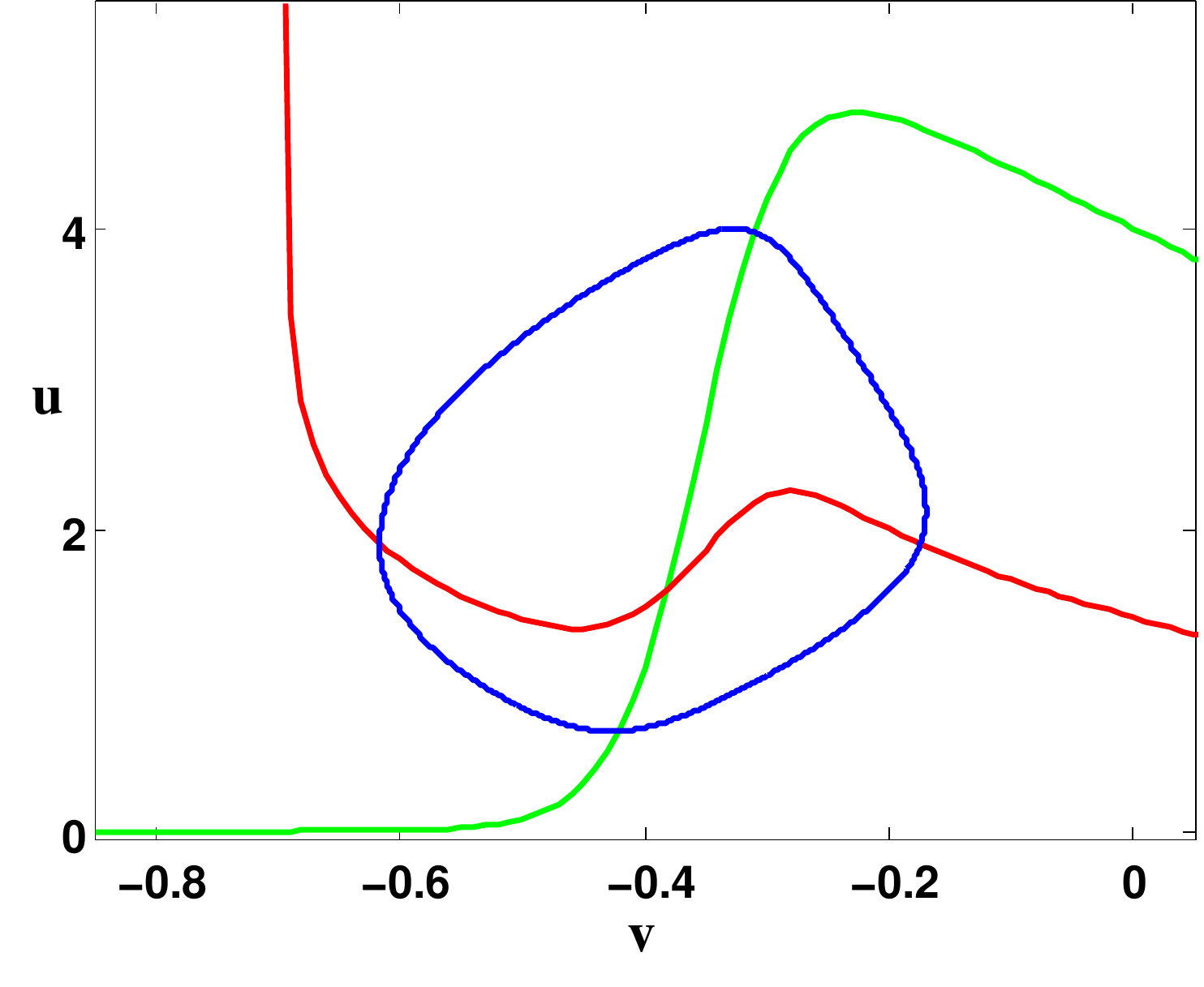, height=2.0in, width=2.5in, angle=0}
{\bf b}\epsfig{figure=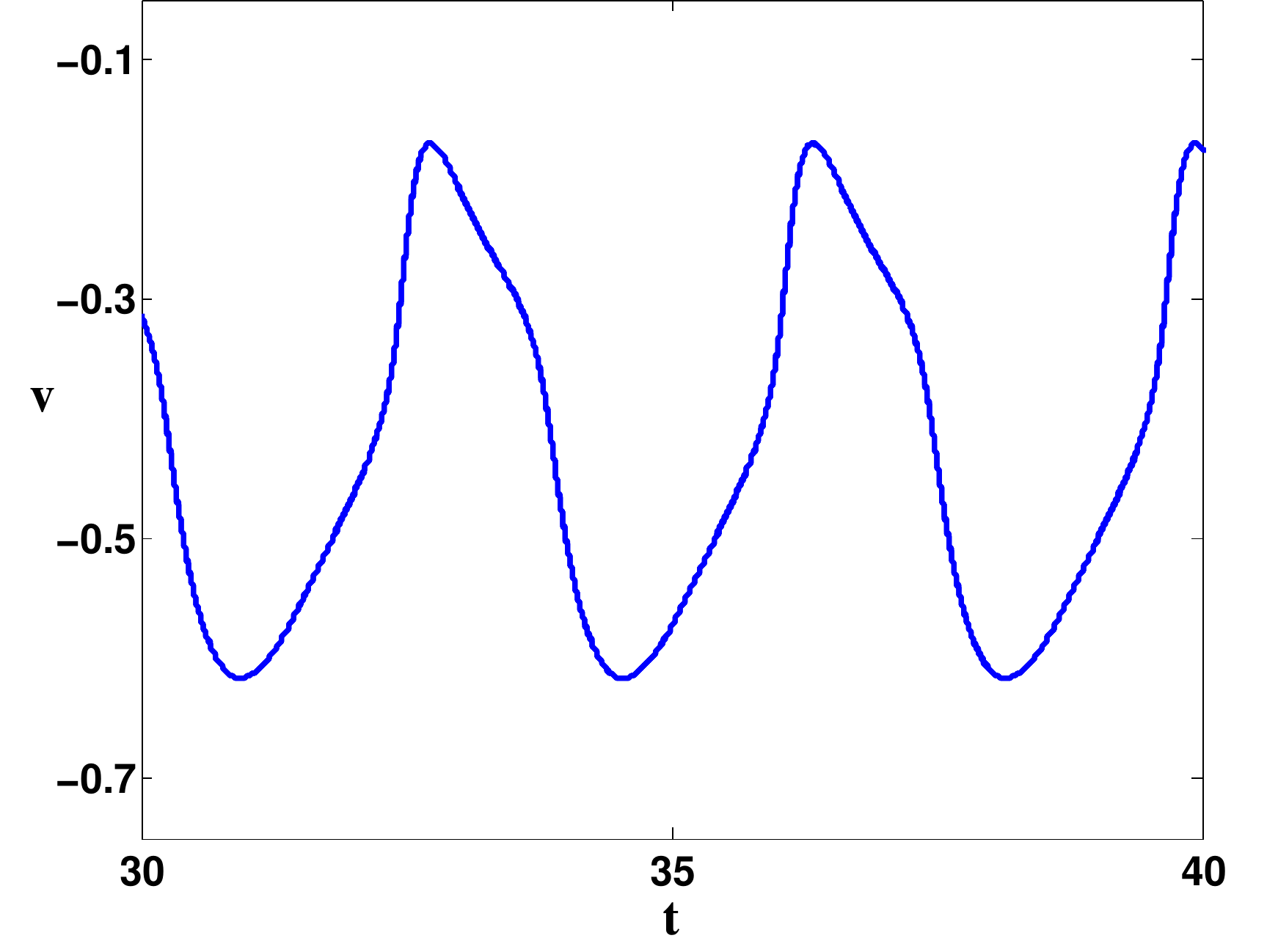, height=2.0in, width=2.5in, angle=0}
{\bf c}\epsfig{figure=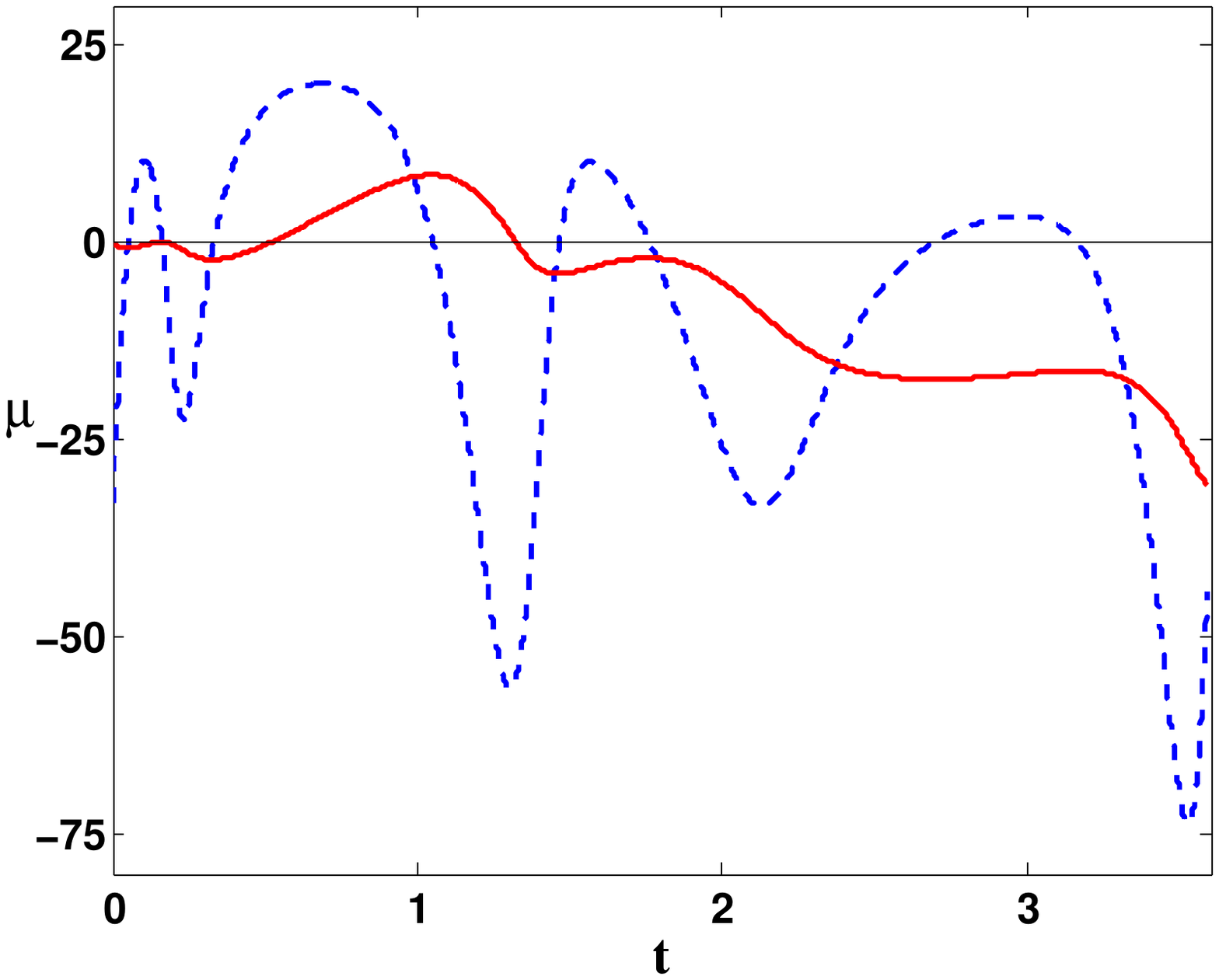, height=2.0in, width=2.5in, angle=0}
{\bf d}\epsfig{figure=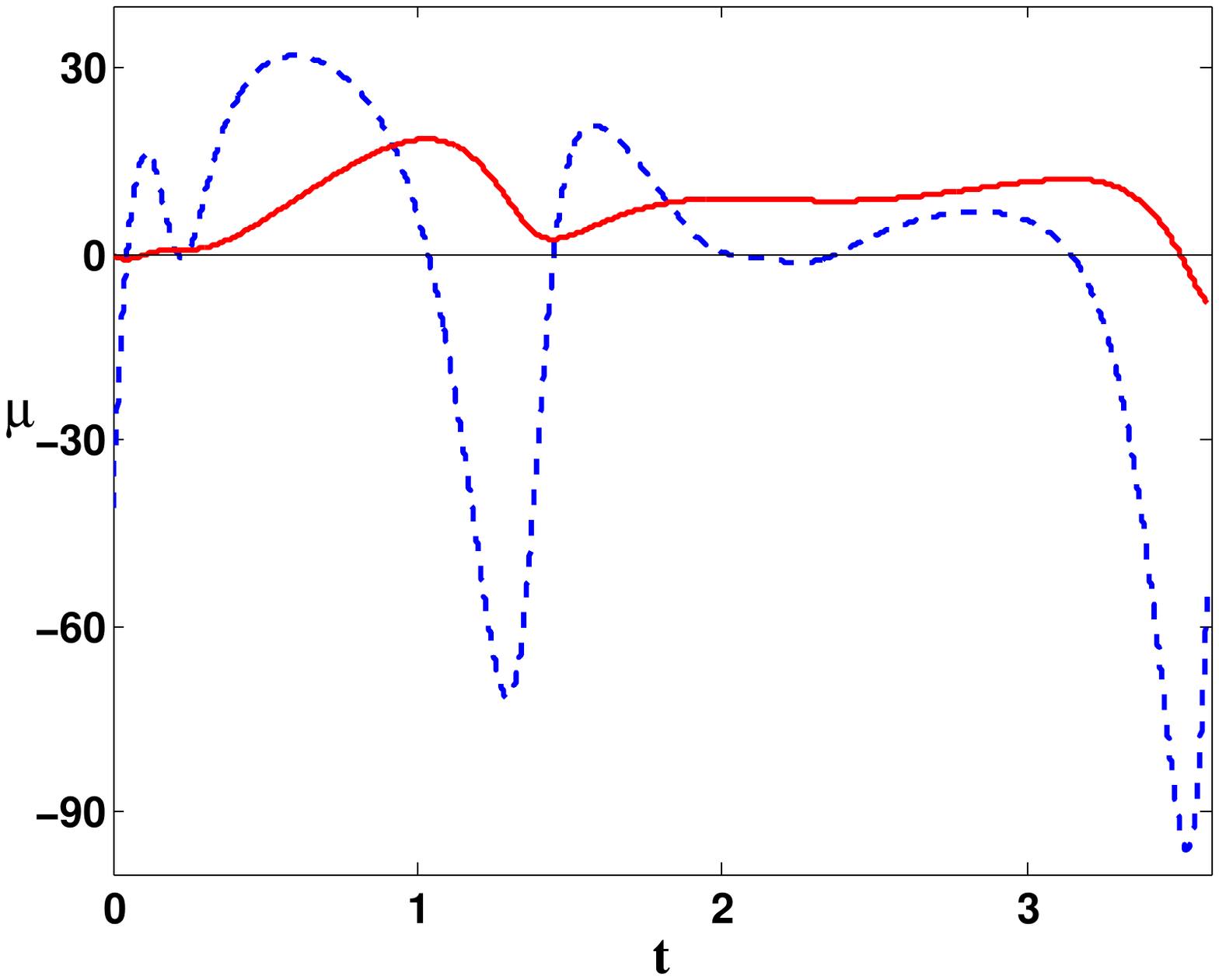, height=2.0in, width=2.5in, angle=0}
\end{center}
\caption{Numerical simulations of compartmental model (\ref{da1})-(\ref{da2}).
(a) Phase plane plot of the limit cycle of the local system.
trajectories of five uncoupled oscillators are plotted for 
(b) Timeseries of five coupled oscillators. 
(c) The largest eigenvalue $\mu(t)$ (dashed line) and
integral $\int_0^t\mu(\theta) d\theta$ are plotted over one
period of oscillations. The eigenvalue takes both positive
and negative values; but since the integral over one
period is negative, the local limit cycle is exponentially stable.
(d)The largest eigenvalue $\mathsf{\lambda_1(t)}$ (dashed line) of matrix
$\mathsf{G}(t)$ determining stability of the synchronous 
regime when the coupling is partial (cf. (\ref{matrixG})).
In solid line we plot $\int_0^t\mathsf{\lambda_1(t) dt}$ for
one period of oscillations. The integral of $\mathsf{\lambda_1(t)}$
over one period is negative. Therefore, the synchronous solution
is stable. 
}\label{f.2}
\end{figure}

We start with the conditions on the coupling matrix $\mathbf{D}$. 
To be specific, we assume the nearest-neighbor coupling
(see Fig.~\ref{f.1}a), i.e., $\mathbf{D}=-\mathbf{\Lambda_0}$. Then 
$$
\mathbf{\hat D}=
\left(\begin{array}{cccccc}
-2 & 1 & 0 & \dots & 0 & 0\\
1& -2& 1& \dots& 0& 0\\
\dots& \dots& \dots& \dots &\dots& \dots \\
0& 0& 0& \dots& 1& -2
\end{array}
\right)\in \Re^{(N-1)\times (N-1)}.
$$
Clearly, $\mathbf{D}\in\mathcal{D}$ because $\mathbf{\hat D}$ is
(symmetric) negative definite. The conditions on $\mathsf{L}$
for both full and partial coupling cases obviously hold.
As is typical for conductance-based models of neurons, away from
the singular limit $\epsilon\to 0$, the analytical 
estimates on the eigenvalues of the variational equations such as 
in Assumptions~\ref{localcycle} and \ref{assumptions}, may be difficult
to derive. We verify (\ref{unistable}) and (\ref{intlambda})
numerically, after we briefly review basic numerics for (\ref{da1})
and (\ref{da2}).
Fig.~\ref{f.2}a shows the limit cycle
of a single uncoupled oscillator in (\ref{da1}) and (\ref{da2}).
The corresponding time series are shown in Fig.~\ref{f.2}b.
In Fig.~\ref{f.2}c, we plot $\mu_1(\theta)$, the largest eigenvalue
of $\mathsf{A^s(\theta)}$. 
In this example $\mathsf{A(\theta)}$ is scalar.
% The next plot in Fig.~\ref{f.2}c illustrates a common albeit little
% commented in the literature fact: stable limit cycles generated
% by neuronal models often are not normally hyperbolic.
Note that over one cycle of oscillations $\mu_1(\theta)$ takes
both positive and negative values. 
However,
since the integral over one cycle of oscillations $\int_0^{\mbox{period}}\mu_1(\theta)d\theta$
is negative (Fig.~\ref{f.2}c), the limit cycle is exponentially stable 
(cf. Lemma~\ref{expstab}).
Therefore, for the fully coupled variant of (\ref{da1}) and (\ref{da2})
all conditions of Theorem~\ref{thm.full} hold.
Note that in the full coupling case, synchronization is determined solely
by the properties of the coupling operator. The only information
about the local system which we use is the exponential stability
of the limit cycle. Conditions for synchronization in partially coupled
systems use the information about the local dynamics
(through matrix $\mathsf{L_1}$ (cf. (\ref{B1s}))) and about the 
coupling operator (through $\ker~\mathsf{L}$). 
% Note that condition 
%(\ref{intlambda}) does not follow from the asymptotic stability 
% of the local limit cycle alone. In fact, the proof of Theorem~\ref{thm.full}
%suggests that the exponential stability of the local limit cycles
% and dissipative coupling may not be sufficient for synchronization
% if the coupling is partial. 
The numerical verification 
of (\ref{intlambda}) for partially coupled variant of (\ref{da1}) 
and (\ref{da2}) is given in Fig.~\ref{f.2}d.

\section{The proof of Theorem~\ref{thm.full}}\lbl{proof}
\setcounter{equation}{0}
The proof proceeds as follows. In \S\ref{coordinates}, we construct
a suitable system of coordinates near the periodic orbit of the
coupled system. In \S\ref{linearized}, we analyze the linear part
of the variational equation. In \S\ref{endgame}
we extend the stability analysis to the full system. 

\subsection{The local coordinates for the coupled system near the limit cycle}
\lbl{coordinates}

The moving coordinates for the coupled system are obtained by combining the 
coordinates (\ref{localvar}) for local systems
\be\lbl{coupvar}
\Re^{Nn}\ni x=\left(
\begin{array}{c}
\mathsf{ x^{(1)}}\\ \mathsf{ x^{(2)}}\\ \vdots \\\mathsf{ x^{(N)}}
\end{array}
\right)=\left(
\begin{array}{c}
\mathsf{ u(\theta^{(1)})+Z(\theta^{(1)})\rho^{(1)}}\\
\mathsf{ u(\theta^{(2)})+Z(\theta^{(2)})\rho^{(2)}}\\
\vdots\\
\mathsf{ u(\theta^{(N)})+Z(\theta^{(N)})\rho^{(N)}}
\end{array}
\right)\mapsto (\theta, \rho)\in (\mathsf{S^1})^N\times\Re^{(N-1)n}.
\ee
where $\theta=(\mathsf{\theta^{(1)},\theta^{(2)}, \dots\theta^{(N)}}),\;
\rho=(\mathsf{\rho^{(1)},\rho^{(2)},\dots,\rho^{(N)}})$.
By following the steps of the proof of Lemma~\ref{lem.polar}, 
for each local system we have
\begin{eqnarray}
\lbl{thetai}
\mathsf{
\dot\theta^{(i)}} &=& \mathsf{1+a^T(\theta^{(i)})\rho^{(i)}+
{v^T(\theta^{(i)})\over |f(\theta^{(i)})|}~
\mbox{CT}+O(|\rho^{(i)}|^2),
}\\
\lbl{rhoi}
\mathsf{
\dot\rho^{(i)}} &=& 
\mathsf{A(\theta^{(i)})\rho^{(i)}+Z^T(\theta^{(i)})~\mbox{CT}+O(|\rho^{(i)}|^2),\; i=1,2,\dots, N,}
\end{eqnarray}
where CT stands for the coupling terms
\be\lbl{CT}
\mbox{CT}:=g\sum_{j=1}^N \mathbf{d_{ij}}\mathsf{L\left( u(\theta^{(i)})-u(\theta^{(j)})\right)}
+g\sum_{j=1}^N \mathbf{d_{ij}}\mathsf{L\left( Z(\theta^{(i)})\rho^{(i)}-Z(\theta^{(j)})\rho^{(j)}\right)}.
\ee
In the moving coordinate frame, the coupling operator becomes nonlinear. We linearize it using the
Taylor's formula. It will be convenient to have Taylor's coefficients appearing in the expansions
for local systems evaluated at a certain common value. For this purpose, we use the average phase
defined by:
\be\lbl{vartheta}
 \vartheta:=\mathbf{\xi}^T\theta \quad\mbox{where}\quad
 \mathbf{\xi}\in\ker\mathbf{D}^T,\;
\sum_{i=1}^N \mathbf{\xi_i}=1.
\ee
The existence of  $\xi\in\Re^N$ with the properties specified 
in (\ref{vartheta}) follows from the following considerations.
First, because
$\rank~\mathbf{D^T}=N-1$, there is nonzero $\mathbf{\xi}\in\ker\mathbf{D}^T$.
$\xi$ can be chosen to satisfy the second condition in (\ref{vartheta})
provided $\mathbf{\xi}^T\mathbf{1_N}\neq 0$. Supose the contrary.
Then, 
$$
\xi\in (\mathbf{1_N})^\perp=(\ker\mathbf{D})^\perp=R(\mathbf{D}^T) \;
\Rightarrow \;\exists\; \eta\neq 0\;:\; \xi=\mathbf{D}^T\eta.
$$ 
Note that $\xi$ and $\eta$ are linearly independent.
Since $\xi\in\ker\mathbf{D}^T$, 
$$
(\mathbf{D}^T)^2\eta=0\;\Rightarrow\; 
\rank~(\mathbf{D}^2)\le N-2.
$$
Therefore, either the geometric multiplicity of the zero eigenvalue
of $\mathbf{D}$ is greater or equal to $2$, or the size of the block 
corresponding zero eigenvalue in the Jordan normal 
form of $\mathbf{D}$ is greater or equal to $2$. 
Either statement contradicts the assumption that $\mathbf{D}\in\mathcal{D}$
(see Lemma~2.5 \cite{medvedev10b}).

Similarly, we define
\be\lbl{varrho}
 \varrho:=(\mathbf{\xi}\otimes\mathsf{I_{n-1}})^T\rho.
\ee
From the definitions of $\vartheta$ and $\varrho$, we have
\begin{eqnarray}\lbl{deviation}
\mathsf{
\theta^{(i)}-\vartheta}&=& 
\mathsf{\sum_{j=1}^N\mathbf{\xi_j}\theta^{(i)}-\sum_{j=1}^N\mathbf{\xi_j}\theta^{(j)}
=\sum_{j=1}^N\mathbf{\xi_j}(\theta^{(i)}-\theta^{(j)})=O(|\phi|),
}\\
\mathsf{
\rho^{(i)}-\varrho}&=&\mathsf{
\sum_{j=1}^N\mathbf{\xi_j}(\rho^{(i)}-\rho^{(j)})=O}(|r|), 
\end{eqnarray}
where $\phi=\mathbf{S}\theta$, $r=(\mathbf{S}\otimes\mathsf{I_n})\rho$,
and $\mathbf{S}$ is defined in (\ref{defineS}).  
 
Using (\ref{deviation}), we represent
\begin{eqnarray}
\lbl{taylor1}
\mathsf{ u(\theta^{(j)})} &=&
\mathsf{ u(\vartheta)+u^\prime(\vartheta)(\theta^{(j)}-\vartheta)+O(|\phi|^2),}\\
\lbl{taylor2}
\mathsf{ Z(\theta^{(j)})} &=&
\mathsf{ Z(\vartheta)+Z^\prime(\vartheta)(\theta^{(j)}-\vartheta)+O(|\phi|^2).}
\end{eqnarray}
By plugging (\ref{taylor1}) and (\ref{taylor2}) into (\ref{CT}) and using (\ref{deviation}), we have
\be\lbl{CT1}
\mbox{CT}=g\sum_{j=1}^N \mathbf{d_{ij}}\mathsf{Lu^\prime(\vartheta)
\left(\theta^{(i)}-\theta^{(j)} \right) }
+g\sum_{j=1}^N \mathbf{d_{ij}}\mathsf{L Z(\vartheta)\left(\rho^{(i)}-\rho^{(j)}\right)}
+O(g|\phi|^2, g|\rho||\phi|).
\ee
Using (\ref{CT1}), we rewrite (\ref{thetai}) and (\ref{rhoi})
\begin{eqnarray}
\lbl{thetai-1}
\mathsf{
\dot\theta^{(i)}} &=& \mathsf{1+a^T(\vartheta)\rho^{(i)}}+
g\sum_{j=1}^N \mathbf{d_{ij}}\mathsf{m(\vartheta)\theta^{(j)}}
+{g\over |f|}\sum_{j=1}^N \mathbf{d_{ij}}\mathsf{U_{ij}(\vartheta)\rho^{(j)}}
+O(|\rho|^2, g|\phi|^2, g|\rho||\phi|),
\\
\lbl{rhoi-1}
\mathsf{
\dot\rho^{(i)}} &=& \mathsf{A(\vartheta)\rho^{(i)}}
+g|f|\sum_{j=1}^N \mathbf{d_{ij}}\mathsf{U^T_{ij}(\vartheta)\theta^{(j)}}
+g\sum_{j=1}^N \mathbf{d_{ij}}\mathsf{M_{ij}(\vartheta)\rho^{(j)}}
+O(|\rho|^2, g|\phi|^2, g|\rho||\phi|),
\end{eqnarray}
where
\be\lbl{mUM}
\mathsf{
m(\vartheta)=v^T(\vartheta)Lv(\vartheta), \; U(\vartheta)=v^T(\vartheta)LZ(\vartheta),} \;\mbox{and}\;
\mathsf{ M(\vartheta)=Z^T(\vartheta)LZ(\vartheta). }
\ee
The averaged system is obtained by multiplying equations for $\theta^{(i)}$ and $\rho^{(i)}$
by $\mathbf{\xi_i}$ and adding them up:
\begin{eqnarray}
\lbl{theta0}
\mathsf{
\dot\vartheta} &=& \mathsf{1+a^T(\vartheta)\varrho}
+O(|\rho|^2, g|\phi|^2, g|\rho||\phi|),
\\
\lbl{rho0}
\mathsf{
\dot\varrho} &=& \mathsf{A(\vartheta)\varrho}+ 
O(|\rho|^2, g|\phi|^2, g|\rho||\phi|, g|\varrho||r|).
\end{eqnarray}
Using $\vartheta$ as an independent variable, we rewrite 
(\ref{thetai-1}) and (\ref{rhoi-1}) as follows
\begin{eqnarray}
\lbl{thetai-2}
\mathsf{
{d\theta^{(i)}\over d\vartheta}} &=& \mathsf{1+a^T(\vartheta)(\rho^{(i)}-\varrho)}+
g\sum_{j=1}^N \mathbf{d_{ij}}\mathsf{m(\vartheta)\theta^{(j)}}
+{g\over |f|}\sum_{j=1}^N \mathbf{d_{ij}}\mathsf{U_{ij}(\vartheta)\rho^{(j)}}\\
\nonumber
&+&O(|\rho|^2, g|\phi|^2, g|\rho||\phi|, g|\varrho||r|),
\\
\mathsf{
{d\rho^{(i)}\over d\vartheta} } &=& \mathsf{A(\vartheta)\rho^{(i)}}
+g|f|\sum_{j=1}^N \mathbf{d_{ij}}\mathsf{U^T_{ij}(\vartheta)\theta^{(j)}}
+g\sum_{j=1}^N \mathbf{d_{ij}}\mathsf{M_{ij}(\vartheta)\rho^{(j)}}
\nonumber\\
\lbl{rhoi-2}
&+&O(|\rho|^2, g|\phi|^2, g|\rho||\phi|, g|r||\varrho|),\; i=1,2,\dots, N,
\end{eqnarray}
or in matrix form for $\tilde x:=(\mathsf{
\theta^{(1)},\rho^{(1)},\theta^{(2)},\rho^{(2)},\dots, \theta^{(N)},\rho^{(N)}
})$  
\begin{eqnarray}\nonumber 
{d\tilde x \over d\vartheta} &=&
\one\otimes\left(\begin{array}{c} 1 \\ \mathbf{0_{N-1}}\end{array}\right)+
\left[ 
\mathbf{I_N}\otimes
\left(\begin{array}{cc} 0 &  \mathsf{a^T} \\ 
                        O &  \mathsf{A} 
\end{array}\right)
+g \mathbf{D}\otimes
\left(\begin{array}{cc} \mathsf{m} & \mathsf{|f|^{-1} U} \\ 
                          \mathsf{|f| U^T} & \mathsf{M} \end{array}\right)
\right] \tilde x \\
\lbl{thetarhomatrix}
&+&O(|\rho|^2, g|\phi|^2, g|\rho||\phi|, g|r||\varrho|),
\end{eqnarray}
where $\mathbf{0_{N-1}}:=(0,0,\dots,0)^T\in\Re^{N-1}$.
By multiplying both sides (\ref{thetarhomatrix}) by 
$\mathbf{S}\otimes\mathsf{I_n}$
and recalling $\phi=\mathbf{S}\theta$ and 
$r=(\mathbf{S}\otimes\mathsf{I_{n-1}})\rho$,
we obtain the system for 
$\tilde y:=(\mathsf{
\phi^{(1)},r^{(1)},\phi^{(2)},r^{(2)},\dots, \phi^{(N-1)},r^{(N-1)}
})$:
\begin{eqnarray}\nonumber
{d\tilde y\over d\vartheta} &=&\left[ 
\mathbf{I_N}\otimes
\left(\begin{array}{cc} 0 &  \mathsf{a^T} \\ 
                        O &  \mathsf{A} 
\end{array}\right)
+g \mathbf{\hat D}\otimes
\left(\begin{array}{cc} \mathsf{m} & \mathsf{|f|^{-1} U} \\ 
                          \mathsf{|f| U^T} & \mathsf{M} \end{array}\right)
\right] \tilde y\\
\lbl{phir}
&+& O(|\rho|^2, g|\phi|^2, g|\rho||\phi|, g|r||\varrho|). 
\end{eqnarray}
The final coordinate transformation in this series is used to make
the matrix multiplied by $\mathbf{\hat D}$ in (\ref{phir}) symmetric:
\be\lbl{psiz}
\psi=\sqrt{|f|} \phi \quad \mbox{and}\quad z={r\over \sqrt{|f|}}.
\ee
Note,
\begin{eqnarray} 
\lbl{psidot}
{d\psi\over d\vartheta} &=& \sqrt{|f|} {d\phi\over d\vartheta}+
{1\over 2} (\mathsf{v^T Dfv)}\psi + O(|\psi||\varrho|),\\
\lbl{zdot}
{dz\over d\vartheta} &=& {1\over\sqrt{|f|}} {d z\over d\vartheta}-
{1\over 4} (\mathsf{v^T Dfv)}z + O(|\psi||\varrho|).
\end{eqnarray}
Using (\ref{psiz}), (\ref{psidot}), and (\ref{zdot}) from (\ref{phir})
we obtain
\be\lbl{finalform}
{dy\over d\vartheta}=B(\vartheta)y + Q_1(y,\varrho),\; 
B(\vartheta):=g B_0(\vartheta)+ B_1(\vartheta),
\ee
where 
$y:=(\mathsf{\psi^{(1)},z^{(1)},\psi^{(2)},z^{(2)},\dots, \psi^{(N-1)},z^{(N-1)}
})$, $Q_1(y,\varrho)=O(|\varrho|^2, g|y|^2,  g|y||\varrho|)$ and
\be\lbl{B0B1}
B_0=\mathbf{\hat D}\otimes
\left(\begin{array}{cc} \mathsf{m} & \mathsf{ U} \\ 
                          \mathsf{U^T} & \mathsf{M} \end{array}\right),
\; B_1=\mathbf{I_{N-1}}\otimes
\left(\begin{array}{cc} {1\over 2}c  & 2\mathsf{v^T(Df)Z} \\ 
                          \mathsf{O} & 
\mathsf{A}-{1\over 2}c\mathsf{I_{n-1}} 
\end{array}\right),
\ee
and $c(\vartheta)=\mathsf{v^T(\vartheta)Df(u(\vartheta))v(\vartheta)}$.
We complement (\ref{psiz}) by the equation for $\varrho$
\be\lbl{dvarrho}
{d\mathsf{\varrho}\over d\vartheta } 
= \mathsf{A(\vartheta)\varrho} +Q_2(y,\varrho),\;
Q_2(y,\rho)=O(|\varrho|^2, g|y|^2, g|y||\varrho|).
\ee

\subsection{The linearized system} \lbl{linearized}
In this subsection, we study the linearization of (\ref{finalform}) 
\be\lbl{systB}
\dot y= B(t)y.
\ee
In the following lemma, we prove that $y=0$ is exponentially stable
solution of (\ref{systB}).
\begin{lem}\lbl{exp}
Let $\Phi(t)$ denote a fundamental matrix solution of (\ref{systB})
and $\Phi(t,s):=\Phi(t)\Phi^{-1}(s)$. Then
\be\lbl{expbnd}
|\Phi(t,s)|\le C_2 \exp\{-\lambda(t-s)\}, \; t\ge s
\ee
for certain positive constants $C_2$ and $\lambda$.
\end{lem}

The proof of Lemma~\ref{exp} follows from Lemmas~\ref{matrixM} and 
\ref{perturbedeigenvalues}, which we prove first.

\begin{lem}\lbl{matrixM}
The spectrum of symmetric matrix
\be\lbl{Msf}
\mathsf{
M(\theta)=
\left(\begin{array}{cc} \mathsf{m(\theta)} & \mathsf{ U(\theta)} \\ 
                          \mathsf{U^T(\theta)} & \mathsf{M(\theta)} \end{array}\right)=
\left(\begin{array}{cc} \mathsf{v^T(\theta)Lv(\theta)} & \mathsf{ v^T(\theta)LZ(\theta)} \\ 
                          \mathsf{Z^T(\theta)Lv(\theta)} & \mathsf{Z^T(\theta)LZ(\theta)} 
\end{array}\right)
}
\ee
coincides with that of $\mathsf{L}$. In particular, for every 
$\mathsf{\theta\in S^1}$, $\mathsf{M(\theta)}$ is a positive semidefinite 
matrix whose rank is equal to $\rank(\mathsf{L})$.
\end{lem}
\pf This follows from 
$$
\mathsf{M(\theta)=O^T(\theta) L O(\theta)},
$$
where $\mathsf{O}=\mbox{col}~(\mathsf{v,z_1,z_2,\dots,z_{n-1}})$ is an
orthogonal matrix.\\
\qed

\begin{lem}\lbl{perturbedeigenvalues}
Let $\lambda_1(t)$ denote the largest eigenvalue of 
\be\lbl{Bg}
B^s(t)=gB_0^s(t)+B_1^s(t).
\ee
Then there exists $g_0>0$ and $\bar\lambda>0$ such that 
\be\lbl{stableB}
\int_0^1 \lambda_1(t)dt =-\bar\lambda,\; g\ge g_0.
\ee
\end{lem}
\pf Matrix $B_0^s(t)=\mathbf{\hat D^s}\otimes\mathsf{M}(t)$ is negative semidefinite
for all $t\ge 0$, because
the eigenvalues of $\mathbf{\hat D^s}$ are negative and those of $\mathsf{M}$ are
nonnegative. Denote the (constant) eigenvalues of $B_0^s$ by $\lambda_k^{(0)}$:
$$
0\ge\lambda_1^{(0)}\ge\lambda_2^{(0)}\ge\dots\ge\lambda_{(N-1)n}^{(0)}.
$$
For small $\delta>0$, the eigenvalues of $B^s_0+\delta B_1^s$ perturb smoothly
\cite{gelfand}
\be\lbl{pertEV}
\lambda_{k,\delta} (t)=\lambda^{(0)}_k+\delta\lambda^{(1)}_k(t)+O(\delta^2),\; 
k=1,2,\dots (N-1)n.
\ee
We consider the full rank coupling case first.
If $\mathsf{L}$ is full rank then so is $\mathsf{M}$. 
Denote
$$
\tilde\lambda:=-0.5\lambda_1^{(0)} <0.
$$
Choose $\delta_0>0$ such that 
$$
\max_k\max_{t\in[0,1]} \lambda_{k,\delta}(t)\le -\tilde\lambda
\quad\mbox{for}~ 0\le\delta<\delta_0.
$$
Then for $g>g_0:=\delta_0^{-1}$, the eigenvalues of $B^s=gB^s_0+B_1^s$ are
negative and are bounded from zero by $-g\tilde\lambda.$ This shows
(\ref{stableB}) for the full coupling case.

It remains to analyze the case $\ker~\mathsf{L}=l>0$, i.e., 
$$
\lambda_1^{(0)}=\dots=\lambda^{(0)}_{l(N-1)}=0,\quad\mbox{and}\quad
\lambda^{(0)}_{l(N-1)+1} =:2\tilde\lambda <0.
$$
By (\ref{pertEV}),
\be\lbl{firstEV}
\lambda_{k,\delta} (t)=\delta\lambda^{(1)}_k(t)+O(\delta^2),\; 
k=1,2,\dots (N-1)l.
\ee
For small $\delta>0$, we have
$$
\int_0^1 \max_{k\in\{1,2,\dots,(N-1)n\}} \lambda_{k,\delta} (t)dt
\le \delta\int_0^1 \max_{k\in\{1,2,\dots,(N-1)l\}} \lambda_{k,\delta} (t)dt +O(\delta^2).
$$ 
Thus, to show (\ref{stableB}) we need to verify 
\be\lbl{intmax}
\int_0^1 \max_{k\in\{1,2,\dots,(N-1)l\}} \lambda_{k,\delta} (t)dt <0.
\ee
For this, we review the construction of the 
correction terms $\lambda^{(1)}_k(t)$
(cf. Appendix \cite{gelfand}). Choose an orthonormal basis for $\ker B_0^s(t)$
$\{\eta_1(t),\eta_2(t),\dots\eta_{l(N-1)}(t)\}$.
Then $\lambda_k^{(1)}(t)$
are the eigenvalues of 
$G=(g_{ij}),$ $g_{ij}=(B_1^s(t)\eta_i(t),\eta_j(t))$. Below we show 
that Assumption~\ref{assumptions} guarantees (\ref{intmax}).
To this end, we construct a basis for $\ker B_1^s(t)$.
Recall that $\{\mathsf{p_1,p_2,\dots,p_l}\}$ stands for the orthonormal
basis of $\ker~\mathsf{L}$ and $\mathsf{O}=\mbox{col}(\mathsf{v,z_1,\dots, z_{n-1}})$.
We choose
\be\lbl{chooseeta}
\eta_i(t)=\mathbf{e_{i_1}}\otimes\mathsf{\xi_{i_2}(t)}, \; i=(i_2-1)l+i_1,\;
i_1\in\{1,2,\dots, N-1\},\;
i_2\in\{1,2,\dots,l\},
\ee
where $\mathsf{\xi_i(t)}=\mathsf{O^T(t)p_i},$
$\mathbf{e_j}= (\delta_{1}^j,\dots,\delta^j_{(N-1)})^T\in\Re^{N-1},$
and $\delta_{i}^j$ denotes the Kronecker delta. 
Vectors in (\ref{chooseeta}) form an orthonormal basis of
$\ker B_1^s(t)$. Further,
\begin{eqnarray*}
g_{ij}(t)&=&(B_1^s(t)\eta_i(t),\eta_j(t))= ((\mathbf{I_{N-1}}\otimes\mathsf{L^s_1})
(\mathbf{e_{i_1}}\otimes\mathsf{\xi_{i_2}(t)}),\mathbf{e_{j_1}}\otimes\mathsf{\xi_{j_2}(t)})=
(\mathbf{e_{i_1}}\otimes\mathsf{L_1^s\xi_{i_2}(t)},\mathbf{e_{j_1}}\otimes\mathsf{\xi_{j_2}(t)})\\
&=& \left\{
\begin{array}{cc}
(\mathsf{L_1^s\xi_{i_2}(t)},\mathsf{\xi_{j_2}(t)})=
(\mathsf{L_1^s O^T(t)p_{i_2}},\mathsf{O^T(t)p_{j_2}}), & \mathsf{i_1=j_1},\\
0, & \mbox{otherwise}.
\end{array}
\right.
\end{eqnarray*}
Thus, $G=\mathbf{I_{N-1}}\otimes\mathsf{G}$ where $\mathsf{G}$ is
defined in (\ref{matrixG}). The eigenvalues of $G$ are those
of $\mathsf{G}$ taken with multiplicity $(N-1)$.
Clearly, Assumption~\ref{assumptions} is equivalent to (\ref{intmax}).
This concludes the proof of the lemma. \\
\qed

\pf (Lemma~\ref{exp})
The statement of Lemma~\ref{exp} follows from 
Lemma~\ref{perturbedeigenvalues} and Lemma~\ref{expstab}.\\
\qed
% \begin{rem}
% As follows from the proof of Lemma~\ref{exp}, the
% largest Floquet exponent associated with (\ref{systB}) 
% is $O(g)$ if the coupling is full and $O(1)$ otherwise.
% \end{rem}

\subsection{The endgame} \lbl{endgame}

To complete the proof of stability of the synchronous solution, we study 
the initial value problem for (\ref{finalform})
and (\ref{dvarrho})
\begin{eqnarray}\lbl{ivp}
\dot x &=& \mathsf{diag}(B(t), \mathsf{A}(t))x +Q(x),\\
\lbl{ic}
x(0) &=& x_0,
\end{eqnarray}
where by abusing notation we denote $x:=(y,\varrho)^T$
and the independent variable by $t$.
Matrices $\mathsf{A}(t)$ and $B(t)$ are defined in (\ref{Atheta})
and (\ref{finalform}).
The nonlinear terms are collected in 
$Q(x):=(Q_1(x), Q_2 (x))= O(g|x|^2)$. We suppress the
dependence of $Q$ on $g$, because once it is chosen sufficiently large
$g$ will be  considered fixed.
Let $X(t)$ denote a principal matrix solution of the homogeneous 
system
$$
\dot x=\mathsf{diag}(B(t), \mathsf{A}(t))x.
$$
By Lemmas~\ref{expstab} and \ref{exp}, for $X(t,s)=X(t)X^{-1}(s)$
we have
\be\lbl{kappa}
|X(t,s)| \le C_3 \exp\{-\kappa (t-s)\},\; t-s\ge 0,
\ee
 for some $C_3>0$ and $0<\kappa<\min\{\mu,\lambda\}$ 
(cf. (\ref{estfund}) and (\ref{expbnd})).

\begin{lem}\lbl{success}
Let $g\ge g_0$ be fixed. Then for any sufficiently small $\epsilon>0$ 
(possibly depending on $g$) and any initial data
\be\lbl{smallic}
|x_0|\le \epsilon
\ee
solution of the initial value problem (\ref{ivp}) and (\ref{ic})
satisfies
\be\lbl{stability}  
\sup_{t\ge 0}|x(t)|\le \epsilon \exp\left\{-{\kappa\over 2} t\right\}.
\ee
\end{lem}
As in the famous theorem of Lyapunov on stability by the linear
approximation, 
the statement of Lemma~\ref{success} follows from the stability 
of the linear part of (\ref{ivp}) captured by (\ref{kappa}).
The proof of the lemma relies on a weaker statement, which we
prove first.

\begin{lem}\lbl{bounded}
Under the assumptions of Lemma~\ref{success}, we have
\be\lbl{stability}  
\sup_{t\ge 0}|x(t)|\le 2\epsilon .
\ee
\end{lem}
\pf 
Fix $\delta$ such that 
\be\lbl{choosedelta}
0<\delta <\min\left\{ {1\over 2\kappa}, {\kappa\over 2}\right\}.
\ee
Because
\be\lbl{nearzero}
Q(0)=0\quad\mbox{and}\quad
{\partial Q(0)\over\partial x}=0,
\ee
for sufficiently small $\epsilon>0$ 
we have
\begin{eqnarray}\lbl{LipQ}
|Q(x)|\le \delta|x|, & |x|\le 2\epsilon,\\
\lbl{LipQa}
|Q(x_2)-Q(x_1)|\le \delta|x_2-x_1|, & |x_{1,2}|\le 2\epsilon.
\end{eqnarray}
Consider a functional sequence defined by
\begin{eqnarray}\lbl{first}
x_1(t)&\equiv& x_0,\\
\lbl{next}
x_{n+1}(t) &=& X(t)x_0 + \int_0^t X(t,s) Q(x_n(s)) ds,\quad n=1,2,\dots .
\end{eqnarray}
We use induction to show that 
\be\lbl{supxn}
\sup_{t\ge 0} |x_n(t)|\le 2\epsilon, \quad n=1,2,\dots .
\ee
The induction hypothesis is verified, using (\ref{kappa}), (\ref{smallic}), 
(\ref{choosedelta}), and (\ref{LipQ}):
$$
|x_2(t)| \le \epsilon\exp\{-\kappa t\} +\delta\epsilon 
\int_0^t\exp\{-\kappa (t-s)\} ds\le 2\epsilon.
$$
Similarly, one shows that (\ref{supxn}) for $n=k$ implies
(\ref{supxn}) for $n=k+1$. Thus, (\ref{supxn}) holds for all
natural $n$.

We complete the proof by showing that $x_n(t)$ uniformly converges
the solution of (\ref{ivp}) and (\ref{ic}). To this end,
we show
\begin{equation}\lbl{geometric}
\sup_{t\ge 0} |x_{k+1}(t)-x_k(t)| \le {1\over 2} \sup_{t\ge 0} |x_k(t)-x_{k-1}(t)|,
\;\; k=2,3,\dots .
\end{equation}
Indeed, by subtracting (\ref{next}) with $n=k$
from (\ref{next}) with $n=k+1$ and using (\ref{kappa}),
(\ref{LipQa}), and (\ref{supxn}),  we have
\begin{eqnarray*}
\sup_{t\ge 0} |x_{k+1}(t)-x_k(t)| &\le& \delta\int_0^t |X(t,s)| 
\sup_{t\ge 0} |x_k(t)-x_{k-1}(t)| ds \le {\delta\over \kappa}
\sup_{t\ge 0} |x_k(t)-x_{k-1}(t)|\\
&\le& {1\over 2}\sup_{t\ge 0} |x_k(t)-x_{k-1}(t)|.
\end{eqnarray*}
Next, consider
\be\lbl{series}
\sum_{k=1}^\infty \sup_{t\ge 0} | x_{k+1}(t)-x_k(t)|.
\ee
By (\ref{geometric}), (\ref{series}) is majorized by the geometric
series $\sum_{k=1}^\infty 2^{-k}$. Therefore, $\{x_n(t)\}$ 
converges uniformly to $x(t)$, the unique solution of (\ref{ivp}),
(\ref{ic}). By (\ref{supxn}), the latter (as the limit of   
$\{x_n(t)\}$) is bounded by $2\epsilon$.
This completes the proof. \\
\qed

\pf (Lemma~\ref{success}) We continue to use the notation introduced
in the proof of Lemma~\ref{bounded}. In particular, positive $\epsilon$ 
and $\delta$ are as chosen above. By the variation of constants formula,
we express the solution of (\ref{ivp}) and (\ref{ic}) as 
\be\lbl{constants}
 x(t)=X(t)x_0+\int_0^t X(t,s) Q(x(s))ds.
\ee
Using (\ref{kappa}), (\ref{smallic}), and (\ref{LipQ}) 
from (\ref{constants}) we have
\be\lbl{preGronwall}
|x(t)|\le \exp\{-\kappa t\}\epsilon +\delta \int_0^t \exp\{-\kappa (t-s)\} |x(s)|ds.
\ee
Rewrite (\ref{preGronwall}) for $y(t):=|x(t)|\exp\{\kappa t\}$:
$$
y(t)\le \epsilon +\delta \int_0^t y(s)ds.
$$
By Gronwall's inequality,
$$
y(t)\le \epsilon\exp\{\delta t\},
$$
and, by recalling the definition of $y(t)$ and (\ref{choosedelta}),
we finally derive
$$
|x(t)|\le \epsilon \exp\{(\delta-\kappa)t\}\le\epsilon 
\exp\left\{{-\kappa\over 2}t\right\}.
$$
\qed  

\vskip 0.25cm
\noindent
{\bf Acknowledgments.} The author thanks 
Kresimir Josic for reading the manuscript 
and providing helpful comments. This work was 
done during sabbatical leave at Program of Applied and 
Computational Mathematics (PACM) at Princeton University.
The author thanks PACM for hospitality.

\renewcommand{\theequation}{A.\arabic{equation}}
\section*{Appendix. Parameter values for (\ref{da1}) and (\ref{da2}) } 
\setcounter{equation}{0}
\label{sec:A}

The equations for the local systems in the neural network (\ref{da1}) 
and (\ref{da2}) are adopted from a nondimensional model of a dopamine 
neuron \cite{MC04} (see also \cite{medvedev10}). 
For biophysical background and details of nondimesionalization, 
we refer an interested reader to \cite{MC04}. 
Terms on the right hand side
of the voltage equation (\ref{da1}) model ionic currents:
a calcium current, a calcium dependent potassium current, and 
a small leak current. 
The equation for calcium concentration (\ref{da2}) takes into account
calcium current and calcium efflux due to calcium pump.
The ionic conductances are sigmoid functions of the voltage 
and calcium concentration
\begin{eqnarray*}
g_1(v)&=& {\bar g_1\over 2} \left(1+\tanh \left({v-a_1\over
a_2}\right)\right),\\
g_2(u) &=& {\bar g_2 u^4 \over u^4 +a_3^4}.
\end{eqnarray*} 
Constants $\bar g_{1,2,3}$ and $E_{1,2,3}$ stand for maximal conductances
and reversal potentials of the corresponding ionic currents;
$a_{1,2,3}$ are constants used in the descriptions of activation
of calcium and calcium dependent potassium currents; $\omega$ and $\epsilon$
are certain constants that come up in the process of nondimesionalization
of the conductance based model of a dopamine neuron
(see \cite{MC04} for details). The values of parameters used
in the simulations shown in Figure~\ref{f.2} are summarized in  
the following table.

\begin{center}
{\sc Table}
\end{center}
\begin{tabular}{|r|r||r|r||r|r||r|r||r|r||r|r||r|r|}
\hline
$E_1$               & $1$     & 
$E_2$                & $-0.9$  & 
$E_3$                & $-0.3$   & 
$\bar g_1$           & $0.8 $   &              
$\bar g_2$           & $2   $  & 
$\bar g_3$          & $1   $  & 
$g$                  & 0.3    \\
$a_1$                & $-0.35$   & 
$a_2$                &$ 1.4 \cdot 10^{-2}$&                      
$a_3$                &$ 1.8$ & 
$\epsilon$           &$ 0.1$&  
$\tau$               & $5.0 $   & 
$\omega$ &         $5.0$ &
  &        \\
\hline
\end{tabular}
\vfill
\break

\end{document}